\documentclass[twocolumn]{aastex631}

\usepackage{multirow}
\usepackage{amsmath}

\shorttitle{NIRC2-Pol First Light}
\shortauthors{Lewis et al.}
\graphicspath{{./}{figures/}}

\begin{document}

\title{NIRC2-Pol: First Light of Near-Infrared Polarimetry on Keck II}

\correspondingauthor{Briley Lewis}
\email{blewis@astro.ucla.edu}

\author[0000-0002-8984-4319]{Briley L. Lewis}
\affiliation{Department of Physics, University of California Santa Barbara, Santa Barbara, CA 91306 USA}
\affiliation{Department of Physics and Astronomy, University of California Los Angeles, Los Angeles, CA 90024 USA}

\author[0000-0003-3567-6839]{Manxuan Zhang}
\affiliation{Department of Physics, University of California Santa Barbara, Santa Barbara, CA 91306 USA}

\author[0000-0001-6205-9233]{Maxwell A. Millar-Blanchaer}
\affiliation{Department of Physics, University of California Santa Barbara, Santa Barbara, CA 91306 USA}

\author{Eduardo Marin}
\affiliation{W.M. Keck Observatory, Kamuela, HI 96743 USA}

\author[0000-0002-9242-9052]{Jayke S. Nguyen}
\affiliation{Department of Astronomy, University of California San Diego, La Jolla, CA 92093}

\author{Carlos Alvarez}
\affiliation{W.M. Keck Observatory, Kamuela, HI 96743 USA}

\author[0000-0001-5082-7442]{Jaren N. Ashcraft}
\affiliation{Department of Physics, University of California Santa Barbara, Santa Barbara, CA 91306 USA}

\author[0000-0002-7361-6776]{Mahawa Cisse}
\affiliation{W.M. Keck Observatory, Kamuela, HI 96743 USA}

\author[0009-0003-2357-4058]{Charles-Antoine Claveau}
\affiliation{Department of Astronomy, University of California Berkeley, Berkeley, CA 94720}


\author{Greg Doppmann}
\affiliation{W.M. Keck Observatory, Kamuela, HI 96743 USA}

\author[0000-0002-0176-8973]{Michael P. Fitzgerald}
\affiliation{Department of Physics and Astronomy, University of California Los Angeles, Los Angeles, CA 90024 USA}

\author{Matthew Freeman}
\affiliation{Department of Astronomy, University of California Berkeley, Berkeley, CA 94720}


\author[0009-0002-0405-4826]{Trisha Hammen}
\affiliation{Department of Electrical and Computer Engineering, Colorado State University, Fort Collins, CO 80523 USA}
\affiliation{W.M. Keck Observatory, Kamuela, HI 96743 USA}

\author{Ryan Hersey}
\affiliation{Department of Physics, University of California Santa Barbara, Santa Barbara, CA 91306 USA}

\author[0000-0002-1583-2040]{Nemanja Jovanovic}
\affiliation{California Institute of Technology, Pasadena, CA 91125}


\author{Scott Lilley}
\affiliation{W.M. Keck Observatory, Kamuela, HI 96743 USA}

\author[0000-0001-9611-0009]{Jessica Lu}
\affiliation{Department of Astronomy, University of California, Berkeley, CA 94720}
\affiliation{Space Sciences Laboratory, University of California, Berkeley, CA 94720}

\author[0000-0001-7809-7867]{James E. Lyke}
\affiliation{W.M. Keck Observatory, Kamuela, HI 96743 USA}

\author{Keith Matthews}
\affiliation{California Institute of Technology, Pasadena, CA 91125}


\author{Dimitri Mawet}
\affiliation{California Institute of Technology, Pasadena, CA 91125}
\affiliation{NASA Jet Propulsion Lab, Pasadena, CA 91109}

\author{William Melby}
\affiliation{Wyant College of Optical Sciences, University of Arizona, Tucson, AZ 85721}

\author[0009-0002-8424-1233]{Thomas McIntosh}
\affiliation{Department of Physics, University of California Santa Barbara, Santa Barbara, CA 91306 USA}



\author{Max Service}
\affiliation{W.M. Keck Observatory, Kamuela, HI 96743 USA}


\author[0000-0002-6356-567X]{Jacob Taylor}
\affiliation{W.M. Keck Observatory, Kamuela, HI 96743 USA}




\begin{abstract}
NIRC2, the Near Infrared Camera 2 on the Keck II telescope, was recently upgraded with a new suite of polarimetric observing modes. The new polarimetry modes (referred to as NIRC2-Pol) open up a wide range of new studies, including investigations of exoplanets, the Galactic center, active galactic nuclei, and solar system objects. The new modes enabled by the upgrade span the 1.1 to 4.1 micron range (i.e. \textit{J} through $L^\prime$ bands) and include imaging polarimetry, coronagraphic imaging polarimetry, and spectropolarimetry. NIRC2-Pol is unique, as Keck II is the largest telescope (10 m) on which AO-fed infrared polarimetry capabilities are available, one of few with $L^\prime$ polarimetric imaging, and the only one where there is both a polarimetric mode and a vortex coronagraph. Here, we introduce the design of NIRC2-Pol, its capabilities, and its current operational status. We also present its first on-sky results: the first $L^\prime$ polarimetric images of the AB Aurigae circumstellar disk. These images more clearly reveal the disk's iconic spiral arms than previous $L^\prime$ total intensity imaging. 
\end{abstract}


\section{Introduction} \label{sec:intro}

Polarimetry is a useful technique for studying various astronomical scenes, including objects with strong magnetic fields or where there is light scattering off dust grains. It is particularly useful for high-contrast imaging of circumstellar disks, as it enables polarimetric differential imaging (PDI); light from a disk is polarized due to scattering, whereas starlight is generally unpolarized, creating a natural separation between the two and therefore a way to remove starlight without disk self-subtraction \citep{kuhn2001imaging,follette2023introduction}.

Polarimetric capabilities in the infrared (IR) currently exist with VLT/SPHERE/IRDIS \citep{de2020polarimetric,van2020polarimetric}, Subaru/SCExAO/CHARIS \citep{gj2021full,lawson2021high}, and Gemini/GPI \citep{perrin2010imaging,millar2016gpi,chilcote2020gpi}; however, these three instruments only operate up to $K$ band ($\lambda_c \sim$2.2$\mu$m), excluding thermal IR wavelengths such as $L'$ band ($\lambda_c\sim$3.8$\mu$m), where only Subaru/IRCS is available \citep{terada2018thermal,watanabe2018near}. These instruments have used polarimetry to produce many results of interest, such as: spectropolarimetry of the transition disk HD 34700 A (Subaru/SCExAO/CHARIS, \citet{chen2024multiband}); polarimetric imaging of brown dwarfs, embedded protoplanets, and disks around evolved binaries (VLT/SPHERE/IRDIS, \citet{van2021survey}, \citet{wahhaj2024pds}, and \citet{andrych2023second}); and deep characterization of the HR 4796A debris disk in polarized light (Gemini/GPI, \citet{perrin2015polarimetry} and \citet{arriaga2020multiband}). Beyond planetary and stellar astronomy, polarimetry is also a technique of interest for the galactic center and active galactic nuclei, e.g. \citet{nishiyama2008magnetic,blinov2021robopol}. A number of past high-contrast instruments were also capable of polarimetric imaging, but have since been decommissioned, such as Subaru/HiCIAO \citep{hodapp2006design,tanii2012high} and VLT/NACO \citep{witzel2011instrumental,de2024polarimetric,millar2020detection}.

In this work, we present a key addition to the current roster of AO-fed NIR imaging polarimeters on major telescopes: NIRC2 Polarimetry (NIRC2-Pol). NIRC2, the Near Infrared Camera 2 for the W.M. Keck Observatory (WMKO), is a near-infrared imager operating from $\sim$1 to 5 $\mu$m (\textit{Y} through \textit{M} bands) on the 10 meter Keck II telescope on Mauna Kea, positioned behind the telescope’s facility adaptive optics system \citep{van2004performance,wizinowich2006wm,lilley2024keck}. NIRC2 is a vacuum cryogenic instrument at a stable temperature of 50 K, and the optical path contains two filter wheels with pupil masks and both narrowband and broadband filters, a grism slide, slits for spectroscopy, and coronagraphic spots. In addition to traditional coronagraphs, NIRC2 has two vector vortex coronagraph masks, one optimized for $L^\prime$ and \textit{M} bands and another for \textit{K} band \citep{castella2016commissioning, xuan2018characterizing}. 

NIRC2 has now been upgraded to enable a new suite of polarimetric observing modes, including imaging polarimetry, high-contrast coronagraphic imaging polarimetry, and medium- to low-resolution spectropolarimetry. To create polarimetric capabilities in NIRC2, three additional hardware components were added: a half-waveplate (HWP) to modulate the angle of polarization for calibration, a Wollaston prism to split the beam into two orthogonal polarization states, and a field mask to restrict the field of view (FOV) to prevent overlap between the ordinary and extraordinary beams. This upgrade began in 2019 with the installation of the field mask and Wollaston prism in NIRC2, and was recently completed with the installation of an optomechanics setup in August 2025 that can hold one of the available HWPs at a time. There is one HWP suitable for \textit{JHK} bands, and one for $L^\prime$. Commissioning observations were completed during Semester 25B, and the mode recently became available to the community starting in Semester 26B.

NIRC2-Pol is unique, as Keck II is the largest telescope (10 m) on which IR polarimetry capabilities are available. It has a broad wavelength range---spanning \textit{J} through $L^\prime$, $\sim$1.1 to 4 $\mu$m---where polarimetry in $L^\prime$ is a particularly unique feature. $L^\prime$ polarimetry can be useful for breaking degeneracies in infrared circumstellar disk modeling and probes bigger dust grains than are visible in \textit{J}/\textit{H}/\textit{K}; these wavelengths also contain multiple spectral features of interest, such as those of polycyclic aromatic hydrocarbons \citep{Kueny2024ApJ,Honda2022PASJ}. Additionally, a number of other science cases beyond circumstellar disks are enabled by $L^\prime$ polarimetry, such as investigations of the nature of bow shocks and features at the galactic center \citep{buchholz2013k}.

NIRC2-Pol also has access to the multitude of other features already available on NIRC2 and Keck II adaptive optics (AO), including grisms for spectropolarimetry, many options of coronagraphic masks, adaptive optics using both a natural guide star (NGS AO) or a laser guide star (LGS AO), and more. NIRC2-Pol is notably the only polarimeter equipped with a vortex coronagraph, making this an interesting testbed for such observations which may be of substantial interest to upcoming missions in development such as the \textit{Habitable Worlds Observatory} \citep{feinberg2024habitable}. Vortex coronagraphs generally have superior contrasts at small inner working angles \citep{mawet2009vector,haffert2025phase}, and polarimetry enables a number of interesting investigations into exoplanet characterization \citep{vaughan2023chasing,gordon2025polarized}. NIRC2-Pol's role in the current landscape of astronomical polarimeters on large telescopes is summarized in Table \ref{tab:inst}.

\begin{table*}
\centering
\begin{tabular}{|l|c|c|cccc|c|c|}
\hline
\multirow{2}{*}{\textbf{Instrument}} & \multirow{2}{*}{\textbf{LGS AO}} & \multirow{2}{*}{\textbf{IR WFS}} & \multicolumn{4}{c|}{\textbf{Wavelengths Covered}}                                                                                      & \multirow{2}{*}{\textbf{Coronagraphs}} & \multirow{2}{*}{\textbf{Telescope Size}} \\ \cline{4-7}
                                     &                                  &                                  & \multicolumn{1}{c|}{\textbf{J}} & \multicolumn{1}{c|}{\textbf{H}} & \multicolumn{1}{c|}{\textbf{K}} & \multicolumn{1}{c|}{\textbf{L'}} &                                        &                                          \\ \hline
VLT/SPHERE/IRDIS                     &                       &                       & \multicolumn{1}{l|}{\checkmark} & \multicolumn{1}{l|}{\checkmark} & \multicolumn{1}{l|}{\checkmark} &                                  & \checkmark                             & 8.2 m                                    \\ \hline
Subaru/SCExAO/CHARIS                 &                       & \checkmark                       & \multicolumn{1}{l|}{\checkmark} & \multicolumn{1}{l|}{\checkmark} & \multicolumn{1}{l|}{\checkmark} &                                  & \checkmark                             & 8.2 m                                    \\ \hline
Gemini/GPI 2.0                           &                      &             & \multicolumn{1}{l|}{\checkmark} & \multicolumn{1}{l|}{\checkmark} & \multicolumn{1}{l|}{\checkmark} &                                  & \checkmark                             & 8.1 m                                    \\ \hline
Subaru/IRCS                          & \checkmark                       & \checkmark                       & \multicolumn{1}{l|}{\checkmark} & \multicolumn{1}{l|}{\checkmark} & \multicolumn{1}{l|}{\checkmark} & \checkmark                       & \checkmark                             & 8.2 m                                    \\ \hline
Keck/NIRC2-Pol                       & \checkmark                       &                      $^*$            & \multicolumn{1}{l|}{\checkmark} & \multicolumn{1}{l|}{\checkmark} & \multicolumn{1}{l|}{\checkmark} & \checkmark                       & \checkmark$^\ddagger$                             & 10 m                                     \\ \hline
\end{tabular}
\caption{Various capabilities for infrared polarimeters available on 8-10 meter class telescopes. NIRC2-Pol is one of only two polarimeters capable of operating in $L'$ band with LGS AO and coronagraphy (alongside IRCS \citep{terada2018thermal}), and is the the only such polarimeter on a 10 meter telescope. 
$^*$IR WFS capabilities are in progress as part of an AO upgrade project at WMKO \citep{lilley2024keck}. $^\ddagger$NIRC2-Pol is the only polarimeter equipped with a \textit{vortex} coronagraph.}\label{tab:inst}
\end{table*}

In this letter, we briefly introduce the design of the NIRC2 Polarimetry mode, its capabilities, and its first on-sky results, with the goal of sharing this new mode with the community and encouraging its use for a variety of science cases. 

\section{NIRC2 Polarimetry Mode Design}

NIRC2-Pol is a dual-channel polarimeter; it uses a polarizing beamsplitter to split the incoming light into two orthogonal polarization states and a half-wave plate (HWP) to modulate the angle of polarization. Through cycles of four critical HWP angles (0°, 45°, 22.5°, 67.5°), it is possible to recover the Stokes vector components $I$, $Q$ and $U$ \citep{stokes1851composition} from the two orthogonal polarization states recorded on the detector. These components can be used to measure the degree and angle of linear polarization.

\begin{figure}
    \centering
    \includegraphics[width=0.9\linewidth]{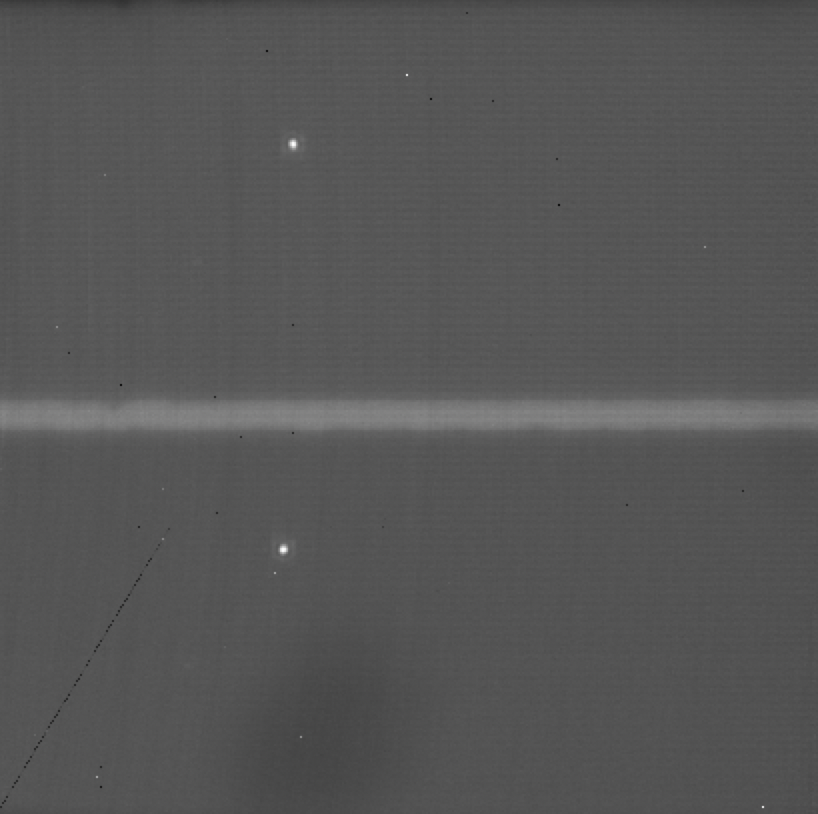}
    \caption{A typical image of a star with the NIRC2-Pol optics in the beam, illustrating how these optics reduce the field-of-view from 10x10" to 5x10". Defects typical of NIRC2 imagery (e.g. the diagonal line, the thumbprint, striping on the detector) are present, along with a bright band in the center of the image. This band is where the fields from the ordinary and extraordinary beams overlap, as the same field is being imaged twice on the same detector in two orthogonal polarization states.}
    \label{fig:flat}
\end{figure}

NIRC2-Pol is enabled by two additional optics in the Keck II AO bench and NIRC2---a HWP and a Wollaston prism---plus a 5x10$^{\prime\prime}$ field mask in NIRC2’s slit and mask sliders to reduce the FOV such that the ordinary and extraordinary beams do not overlap (see Fig. \ref{fig:flat}). The Wollaston prism, a polarizing beamsplitter, is located in the NIRC2 outer filter wheel, and the HWP is located in the Precision Calibration Unit (PCU2), which is at the front (telescope side) of the Keck II AO bench. 

This mode was designed as an upgrade to the existing NIRC2 instrument, as part of a new Precision Calibration Unit analogous to the Pinhole Calibration Unit (PCU1) on Keck I \citep{freeman2023optical,claveau2026inprep}. The PCU2 contains the optomechanics necessary to rotate and translate the HWPs or, alternatively, a pinhole mask that can be used to analyze distortion on the detector for precision astrometry \citep{lin2024design}. The PCU2 unit includes a mount in a rotation stage that can hold one of the following optics at a time: a pinhole mask, a half-wave plate optimized for \textit{JHK}-band polarimetric observations, or a half-wave plate optimized for $L^\prime$ polarimetric observations. To swap between these three, the optic in the PCU2 must be manually switched by WMKO staff. This swap procedure is a limiting factor for observations, and must be accounted for in proposal and observation planning, as the observatory requires significant notice to plan and perform this daytime task.  

\subsection{HWP Specifications}

The \textit{JHK} HWP was fabricated by Bernard Halle and is effective over the wavelength range 1.1--2.4$\mu$m. It is an achromatic half wave retarder made of two plates---one quartz (1.48 mm thickness) and one MgF$_2$ (1.16 mm thickness)---with an air gap and four spacers. The waveplate has a clear aperture of 53 mm diameter. The quartz side of the \textit{JHK} HWP is coated with TelAztek RAR Nano-Texture broadband anti-reflective coating, which does not significantly affect the optic's retardance \citep{melby2024half}. The throughput of the JHK HWP is over 92\% across the wavelength range tested in lab (1.1--1.9$\mu$m), and is predicted to be this high or better at longer wavelengths. The manufacturer specification for the JHK HWP retardance is $\lambda$/2 $\pm$ $\lambda$/70. Lab measurements using a dual-rotating retarder polarimeter (DRRP) show that the deviations of the waveplate from $\lambda$/2 are about $\lambda$/59 in J, $\lambda$/510 in H, and $\lambda$/122 in K \citep{zhang2026inprep}. We note that the \textit{JHK} HWP was damaged in transit to the observatory, and as a result has a small crack on the edge of the plate. This defect is not in the beam path with standard target placement in the center of the field and standard HWP placement in the center of the beam.

The $L^\prime$ waveplate was fabricated by the Karl Lambrecht Corporation and is effective over the wavelength range 3.4--4.1$\mu$m. It is a zero order half-wave plate made with two air spaced MgF$_2$ half wave retarders (2.75 mm thickness) with a clear aperture of 53 mm. Its throughput, as specified by the manufacturer, is $>$90\% across its effective wavelength range. To determine the $L^\prime$ plate's retardance, laboratory measurements were taken using the DRRP between 1-2 microns and extrapolated to $L^\prime$ wavelengths, finding less than $\lambda$/5 variation across L’ band \citep{melby2024half}.


\section{Observing with NIRC2-Pol}

NIRC2-Pol is a non-facility visitor mode on Keck II as of Semester 26B; as of this writing, the mode is selectable on the Keck cover sheet for proposals (abbreviated as \texttt{NIRC2pJHK-NGS}, \texttt{NIRC2pL-NGS}, \texttt{NIRC2pJHK-LGS}, and \texttt{NIRC2pL-LGS}), header keywords are saved for relevant polarimetric quantities (e.g. HWP angle, PCU2 position), and data are automatically ingested into the Keck Observatory Archive (KOA) \citep{berriman2005design,berriman2014design,oluyide2024observers}. A number of tools have been developed to assist an observer with NIRC2-Pol observations, such as observing scripts, an updated calculator for observing overheads including HWP rotation, and a GUI to show the HWP angle, 5x10" mask status, and PCU2 position. The \texttt{ZetaPersei} package provides tools for planning polarimetric standard stars and polarimetric calibration for NIRC2 \citep{zhang2025zetapersei}, and a framework for polarimetric signal-to-noise ratio calculations is available in \citet{nguyen2021exposure}. A full operations guide is also available, with detailed instructions for observing with the new mode \citep{lewis_2026_20737935}.

\subsection{Executing Polarimetric Observations}

To recover the Stokes components $Q$ and $U$ (and therefore the degree and angle of linear polarization), polarimetric observations must be executed in cycles of four critical HWP angles: 0°, 45°, 22.5°, and 67.5.° NIRC2-Pol has an observing script that allows the user to specify a number of HWP cycles to execute, with the option for dithering on the target. Generally, a HWP cycle should aim for a total time of $\lesssim$ 4 minutes (1 minute per angle) to avoid smearing of any time-dependent changes in the polarization signal, and so integration times and coadds should be chosen accordingly. 

There are three modes currently available and tested on-sky with NIRC2-Pol in both \textit{JHK} and $L^\prime$: standard imaging polarimetry, high-contrast imaging polarimetry with traditional Lyot coronagraphs, and high-contrast imaging polarimetry with a vortex coronagraph. A vortex coronagraph has spatially-varying retardance, therefore altering the polarization vector of incoming light and making data processing more complex. A data processing pipeline for NIRC2-Pol---including an effort to incorporate vortex observations and polarimetry---is currently in development \citep{lewis2026inprep}. Low-resolution spectropolarimetry with the NIRC2 grisms is possible, but has not been tested.

Observing in $L^\prime$ also presents challenges due to the substantially higher thermal background. Dome flats without the Wollaston in and the full detector size are not generally possible, as they will saturate the detector. The thermal background changes on quite short timescales and significant effort is required to remove this effect, as described in \citet{nguyen2025ground}. With NIRC2-Pol, we recommend taking sky flats/backgrounds with each target for multiple HWP cycles and dithering during observations with an ABBA pattern whenever possible.

Although flat fielding can be done in the standard way (i.e. no polarimetric optics in the beam), it is recommended to take ``polarimetric flats'' to account for imperfections introduced by the additional optics. These flats should be taken with the Wollaston and HWP in the beam, and the HWP rotating to its standard set of critical angles. 

\subsection{Polarimetric Efficiency}\label{poleff}

\begin{figure}
    \centering
    \includegraphics[width=\linewidth]{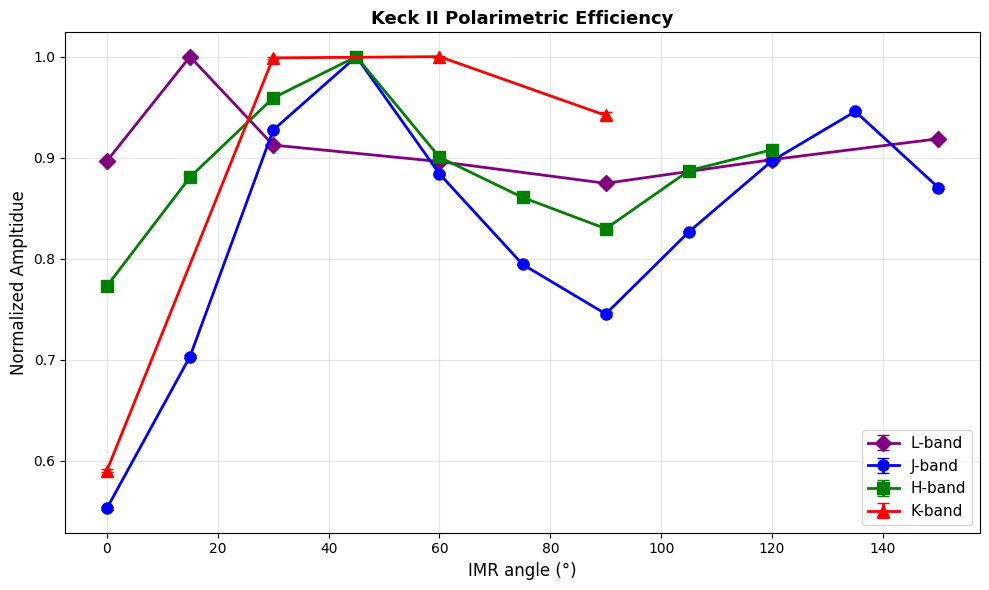}
    \caption{Relative polarimetric efficiency at each wavelength band for NIRC2-Pol. Each curve is normalized to one at its most efficient angle; a determination of the \textit{absolute} efficiency can only be made with a full Mueller matrix model of the system \citep{zhang2026inprep} The wavelength dependence of the IMR's behavior is clear, with a significantly smaller effect at $L^\prime$ band.}
    \label{fig:imr}
\end{figure}

Observers must consider the angle of the image rotator (IMR, a.k.a. K-mirror) during polarimetric observations, as it strongly affects the instrument's polarimetric efficiency. The wavelength-dependent retardance of the IMR can cause crosstalk between linear and circular polarization, reducing the efficiency of the Wollaston prism, which is only sensitive to linear polarization. To determine the most optimal IMR angles (measured relative to its position on the AO bench) to reduce the conversion of linear polarization into circular polarization, we used the internal calibration flats listed in Table \ref{tab:obs} to create Fig. \ref{fig:imr}. For each wavelength and IMR angle, we computed the normalized beam difference $\eta$ using:

\begin{equation}
\begin{aligned}
\eta = \frac{(I_{top} - I_{bottom})}{(I_{top} + I_{bottom})}
\label{Normalized_Difference_Equations}
\end{aligned}
\end{equation}\label{eq:normdiff}

\noindent where $I_{top}$ and $I_{bottom}$ are fluxes from aperture photometry in the top and bottom beams of the Wollaston prism (see Figure \ref{fig:flat} for an example of this splitting). 

We then fit the HWP angle dependence of these normalized differences as
\begin{equation}
\eta(\theta_{\rm HWP}) =
c_0 + c_1{\rm cos}(4\theta_{\rm HWP})
    + c_2{\rm sin}(4\theta_{\rm HWP}) .
\label{eq:imr_hwp_fit}
\end{equation}
The independent ${\rm cos}(4\theta_{\rm HWP})$ and ${\rm sin}(4\theta_{\rm HWP})$ terms allow the modulation to have an arbitrary phase, rather than assuming that the extrema occur at a particular HWP angle. Here, $c_1$ and $c_2$ are the fitted cosine and sine components respectively of the $4\theta_{HWP}$ modulation. Together, $c_1$ and $c_2$ define the modulation amplitude
\begin{equation}
A=\sqrt{(c_1^2+c_2^2)} .
\label{eq:imr_modulation_amplitude}
\end{equation}
The plotted quantity in Fig. \ref{fig:imr} is $A$ normalized by the maximum value of $A$ within each wavelength band.

The polarimetric efficiency of the instrument is most highly affected by the IMR at shorter wavelengths. In \textit{J}  and \textit{H} bands, the angle of the IMR relative to the AO bench that produces the best polarimetric efficiency is 45 degrees as shown in Figure \ref{fig:imr}; angles around 0 and 90 degrees should be avoided when possible, especially for quantitative polarimetry. In \textit{K} band, the same angles should be considered but the effect is less severe. In $L^\prime$ band, the polarimetric efficiency is generally stable regardless of IMR angle. 

\subsection{Effects on Adaptive Optics Performance}

Keck II Adaptive Optics (AO) can be used with NIRC2-Pol in the same manner as standard NIRC2 imaging, with minor impacts on the Strehl ratio. For natural guide star (NGS) AO, from commissioning observations in \textit{JHK} bands before the recent HAKA deformable mirror upgrade \citep{lilley2024keck}, we measured an average Strehl ratio of 56.6$\pm$0.9\% with standard NIRC2 observations and an average Strehl of 52$\pm$2\% with NIRC2-Pol. We note that seeing degraded slightly throughout the course of these observing sequences 
with high cirrus clouds intermittently present, which may explain the slight discrepancy between the average Strehl ratios; however, measurements were taken alternating polarimetric vs. non-polarimetric modes to mitigate the effect of changing seeing on this comparison.

The Strehl ratio degradation is noticeably worse with the insertion of the $L^\prime$ band waveplate. During commissioning, we observed an average Strehl ratio of 40$\pm$3\% with the polarimetric optics, and an average of 47$\pm$4\% without. Removing the HWP did not improve the Strehl ratio, indicating that NCPAs from the Wollaston are a likely culprit for the degradation of the correction. Fast and Furious focal plane wavefront sensing \citep{bos2021fast} works without any modifications, and showed $\sim$5\% improvement to the Strehl ratio in L’ band with NGS AO.

Commissioning tests reveal similar degradation of Strehl in $L^\prime$ band with the polarimetric optics for laser guide star (LGS) AO. We observed an average Strehl of 19$\pm$4\% with the polarimetric optics, and an average of 26$\pm$5\% without. The HWP vignettes the full telescope focal plane, and the effective field of regard is slightly less than 30$^{\prime\prime}$ in radius with the HWP in position, placing constraints on choices of tip/tilt stars. The $L^\prime$ HWP reduces the laser throughput to the WFS camera by approximately 0.2 magnitudes, an amount that should not affect LGS performance. Comparable measurements have not been made for LGS AO with the \textit{JHK} HWP, although similar performance is expected.

\subsection{Polarimetric Calibration}\label{polcal}

A number of calibrations are necessary for polarimetric observations, especially if any quantitative measurements (degree and angle of linear polarization) are desired. Namely, we must determine the fast axis of the HWP and model the instrumental polarization, crosstalk, and polarimetric efficiency of the system. Other calibrations---such as dark subtraction and sky subtraction---are done in the standard manner for NIRC2.

A full Mueller Matrix model of the system is currently in development \citep{zhang2026inprep}; this model will allow us to correct for instrumental polarization and enable high-precision quantitative polarimetry similar to CHARIS and VAMPIRES on Subaru/SCExAO \citep{gj2021full,zhang2023characterizing} and SPHERE/IRDIS on the VLT \citep{van2020polarimetric,de2020polarimetric}. Without the full Mueller Matrix model for instrumental polarization, NIRC2-Pol is still capable of relative and qualitative polarimetry, including polarimetric differential imaging for circumstellar disk morphology \citep{follette2023introduction}.

In this work, 
we estimate the instrumental polarization (i.e. degree of linear polarization induced by the telescope and instrument optics) in $L^\prime$ from observations of two unpolarized standard stars (HD 69699 and HD 65970, both observed on 3 December 2025; Table \ref{tab:obs}) from altitudes of $46^{\circ}$ to $74^{\circ}$. 

Aperture sizes are determined using curve of growth analysis on each frame acquired on the two unpolarized standards after background subtraction. We perform both dither subtraction (where the two dither positions are differenced) and sky subtraction using annuli from aperture photometry, to ensure adequate removal of the highly spatially and temporally variable thermal background. For each complete HWP cycle, we take the double differences (described further below in Section \ref{science}) to retrieve Stokes parameters, then calculate the normalized polarized intensity ($PI = \sqrt{q^2 + u^2})$ where $q = Q/I$ and $u = U/I$. The uncertainty in the polarized intensity is calculated by propagating photometric uncertainties within each beam. 

\begin{figure}
    \centering
    \includegraphics[width=\linewidth]{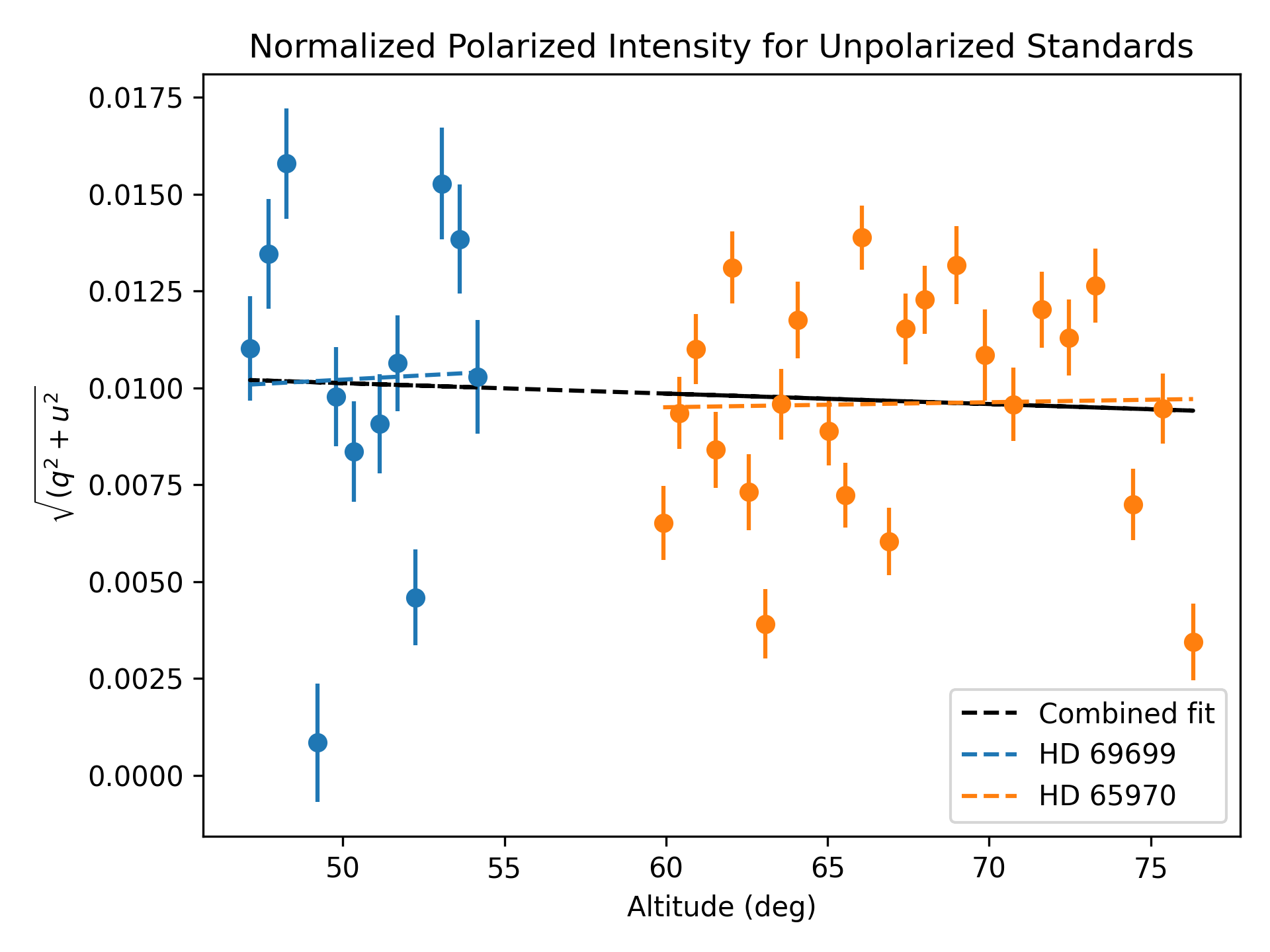}
    \caption{Polarized intensity versus altitude for the two observed unpolarized standards in $L^\prime$. A linear regression to the data shows a near-constant offset across altitudes, which we interpret as a rough estimate of the instrumental polarization in $L^\prime$: 0.98$\pm$0.33\%.}
    \label{fig:unpol_PI}
\end{figure}

Without a Mueller matrix model to account for crosstalk between polarization states downstream of the HWP, $Q$ and $U$ may not be accurate as signal could be exchanged between the two states. However, polarized intensity is not affected by crosstalk between $Q$ and $U$, as it is a combination of these two states. As diattenuation is suppressed downstream of the HWP, observations of unpolarized standard stars therefore theoretically isolate the polarized intensity induced by M3. The polarized intensity versus altitude for the observed unpolarized standards is shown in Figure \ref{fig:unpol_PI}. A linear regression to these data shows that the polarized intensity is constant across altitudes (the slopes of the best fit lines are $<5\times10^{-5}$), but non-zero. Given that we do not see significant modulation with altitude, crosstalk from linear polarization to circular is likely minimal. As a result, this offset is interpreted as a rough estimate of the instrumental polarization (IP) off M3 in $L^\prime$: 1.0$\pm$0.3\%. 

\begin{table*}[]
\centering
\begin{tabular}{|l|l|l|l|l|l|l|l|l|}
\hline
\textbf{Target} & \textbf{Type}         & \textbf{Obs. Date} & \textbf{$\lambda$} & \textbf{$t_{\rm int}$} & \textbf{$t_{\rm exp}$} & \textbf{Coadds} & \textbf{Detector Mode} & \textbf{Dither} \\ \hline
AB Aurigae      & Science               & 7 Dec 2025         & $L^\prime$        & 35.8 min               & 0.45 sec               & 45              & MCDS 4                 & Yes                \\ \hline
HD 69699        & Unpolarized Standard  & 3 Dec 2025         & $L^\prime$        & 35 min                 & 0.7 sec                & 30              & CDS                    & Yes                \\ \hline
HD 65970        & Unpolarized Standard  & 3 Dec 2025         & $L^\prime$        & 28.8 min               & 0.6 sec                & 30              & CDS                    & Yes                \\ \hline
Flat            & Calibration  & 26 Nov 2025        & \textit{J}         & 30 sec                    & 30 sec                 & 1               & CDS                    & No                 \\ \hline
Flat            & Calibration  & 26 Nov 2025        & \textit{H}         & 30 sec                    & 30 sec                 & 1               & CDS                    & No                 \\ \hline
Flat            & Calibration  & 27 Nov 2025        & \textit{K}         & 110 sec                    & 110 sec                & 1               & CDS                    & No                 \\ \hline
Twilight Sky Flats        & Calibration  & 7 Dec 2025         & $L^\prime$        & 15 sec                    & 0.5 sec                & 30              & CDS                    & No                 \\ \hline
\end{tabular}\caption{On-sky observations used in this work, where $t_{\rm int}$ is the total integration time on source in a given configuration and $t_{\rm exp}$ is the individual exposure time.
The unpolarized standard stars are used to derive a rough estimate of the instrumental polarization in Section \ref{polcal}, and the AB Aurigae data are discussed in Section \ref{science}. Ccalibration flats (used to derive the relative polarimetric efficiency, Figure \ref{poleff}) were taken from HWP angles 0 to 90 in 10$^\circ$ increments and IMR angles 0 to 150 in 15$^\circ$ increments for \textit{J}, \textit{H}, and $L^\prime$. Calibration flats were taken from HWP angles 0 to 90 in 10$^\circ$ increments and IMR angles 0 to 90 in 30$^\circ$ increments for \textit{K}. All calibration flats were taken using the dome lamps except for $L^\prime$, where the twilight sky was used. 
}\label{tab:obs}
\end{table*}

\section{First On-Sky Results}\label{science}

\begin{figure*}
    \centering
    \includegraphics[width=0.9\linewidth]{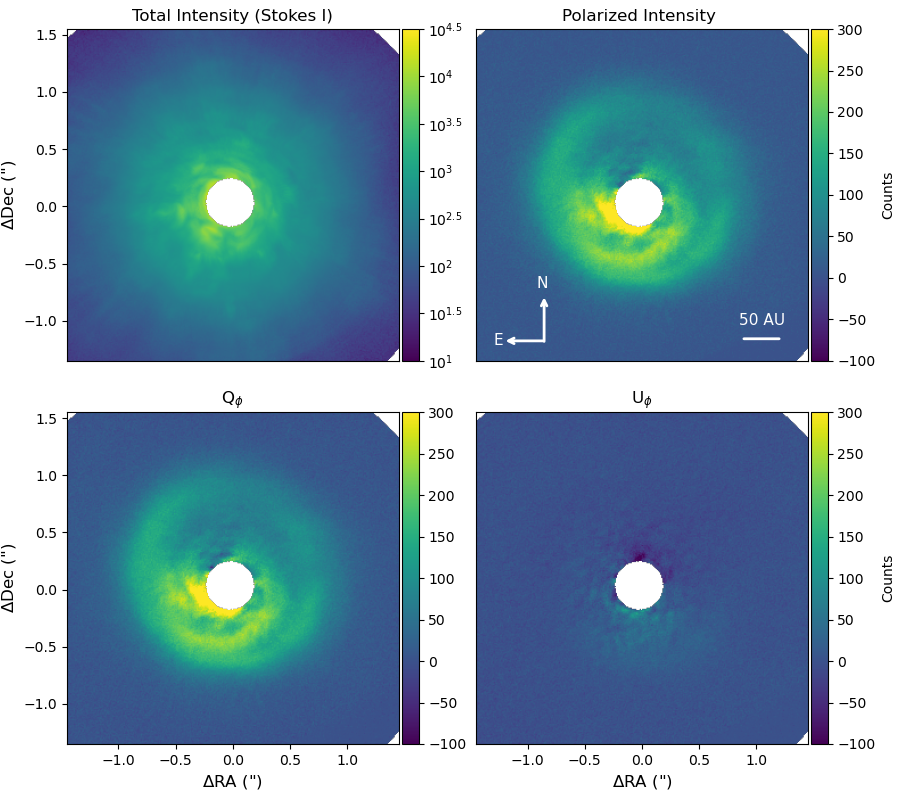}
    \caption{The AB Aur circumstellar disk as seen by NIRC2-Pol during commissioning in $L^\prime$ total intensity (top left; no PSF-subtraction) and polarized intensity (top right; Stokes \textit{Q$_\phi$} and Stokes \textit{U$_\phi$} added in quadrature), as well as $Q_\phi$ (bottom left) and $U\phi$ (bottom right). For a circumstellar disk without multiple scattering, $U_\phi$ should only contain noise.}
    \label{fig:abaur}
\end{figure*}

During commissioning, we observed the well-studied transition disk around AB Aurigae, resolving it for the first time in $L^\prime$ polarized intensity. AB Aur was observed with non-coronagraphic polarimetric imaging on 7 Dec 2025 for a total integration time of 35.8 minutes, with individual exposures of 0.45 seconds and 45 coadds, the detector in MCDS (n$_{\rm samp}$=4) readout mode, the image rotator in vertical angle (\texttt{VERTANG}) mode, and two dither positions. With the chosen exposure times, the core of AB Aur was saturated.

\subsection{Data Processing}

The data were first dark-subtracted and flat-fielded, and bad pixels were cleaned up using median replacement with a 7x7 pixel window. The data were then split into two halves for the ordinary/extraordinary beams, the mean thermal background of each frame was subtracted, and the frames were roughly centered. To retrieve Stokes \textit{Q} and \textit{U} in the instrument frame, we take double differences, where $I_{\rm top}(\theta)$ and $I_{\rm bottom}(\theta)$ are the two orthogonal polarization states on the detector for a given HWP angle $\theta$ and $Q^+$ and $Q^-$ are the single differences:
\begin{equation}
\begin{split}
    Q &= \frac{1}{2}(Q^+ - Q^-) \\
      &= \frac{1}{2}(I_{\rm top}(0^\circ) - I_{\rm bottom}(0^\circ)) \\
      &\quad - (I_{\rm top}(45^\circ) - I_{\rm bottom}(45^\circ))
\end{split}
\end{equation}
\begin{equation}
\begin{split}
    U &= \frac{1}{2}(U^+ - U^-) \\
      &= \frac{1}{2}(I_{\rm top}(22.5^\circ) - I_{\rm bottom}(22.5^\circ)) \\
      &\quad - (I_{\rm top}(67.5^\circ) - I_{\rm bottom}(67.5^\circ))
\end{split}
\end{equation}

Double-differencing in this manner heavily suppresses instrumental polarization downstream of the HWP. We then rotate the polarization state from the instrument frame of reference to the sky frame, using a rotation matrix that accounts for the sky position, image rotator position relative to the bench, and the offset of the HWP fast axis from zero. We derive a fast axis offset value ($\theta_{\rm off}$) from the data themselves by finding the value that best aligns the polarimetric quadrupole pattern; this method finds a value of $\theta_{\rm off} = -13.1^\circ$. We note that this may not be the true fast axis offset; in the absence of the full Mueller matrix model, there is potential for $\theta_{\rm off}$ to absorb effects from unaccounted retardance in the IMR and/or other optics. In terms of the relevant NIRC2 header keywords, the overall polarimetric rotation is then: $\theta_{\rm rot}$ = -2$\times$\texttt{PARANG} + 2$\times$\texttt{EL} + 2$\times$\texttt{ROTPDEST} + 4$\times$$\theta_{\rm off}$, using the following equations to relate the measured $Q$ and $U$ to those in the sky frame $Q'$ and $U'$:
\begin{equation}
    Q' = Q{\rm cos}\theta_{\rm rot} + U{\rm sin}\theta_{\rm rot}
\end{equation}
\begin{equation}
    U' = -Q{\rm sin}\theta_{\rm rot} + U{\rm cos}\theta_{\rm rot}
\end{equation}

After this double-differencing procedure, we finally de-rotate the sky-frame Stokes images $Q'$ and $U'$ frames to North up - East left using \texttt{pyklip.rotate} \citep{wang2015pyklip,Wang2015ascl} and the equation for the derotation angle from \citet{Service2016PASP}.

We then transform to $Q_\phi$ and $U_\phi$ radial Stokes parameters, where $\phi$ is the azimuthal angle around the disk (-$\pi$ to $\pi$ measured counterclockwise from -x); in this frame, $Q_\phi$ contains disk signal and $U_\phi$ should contain noise:
\begin{equation}
    Q_\phi = Q{\rm cos}(2\theta) + U{\rm sin}(2\theta)
\end{equation}
\begin{equation}
    U_\phi = -Q{\rm sin}(2\theta) + U{\rm cos}(2\theta)
\end{equation}

As we have assumed ideal optics in the polarimetric rotation above, instrumental polarization from M3 and crosstalk between polarization channels likely remains in the data. To further improve frame registration/centering and empirically remove instrumental polarization, we perform a minimization of $U_\phi$ on each cycle to find the best fit center positions and two instrumental polarization terms (one each for $Q$ and $U$). To create the final combined images, we perform a weighted combination where the frames are weighted inversely to the noise (i.e. those with the lowest standard deviation in $U_\phi$---the least noise---are weighted the highest). The final polarized intensity image can be seen in Figure \ref{fig:abaur}. $U_\phi$ should only contain noise for a disk without multiple scattering, and the resulting $U_\phi$ images from this reduction contain little signal, giving us confidence in this data reduction and empirical IP removal process. 

Total intensity can be recovered from polarimetric observations by averaging the double sums ($I_Q$ and $I_U$, Equation \ref{doublesum}) derived from the 0/45$^\circ$ and 22.5/67.5$^\circ$ cycles respectively. 
\begin{multline}\label{doublesum}
I_{Q/U} = \frac{1}{2}(I_{Q^+/U^+} + I_{Q^-/U^-}) \\
= \frac{1}{2}(I_{\rm top}(0/22.5^\circ) + I_{\rm bottom}(0/22.5^\circ) \\
+ I_{\rm top}(45/67.5^\circ) + I_{\rm bottom}(45/67.5^\circ))
\end{multline}

The recovered total intensity polarimetric data were also reduced as a typical angular differential imaging dataset (see Figure \ref{fig:spirals}). The two polarization components were sub-pixel centroid aligned prior to adding them together. The full aligned data cube was reduced using principal component analysis (PCA) as implemented in \texttt{ADI.jl} \citep{Lucas2020}. To reduce self-subtraction, the data was filtered by removing frames that had a parallactic rotation of at least 50\textdegree. With this constraint, the highest signal data was produced by subtracting 5 PCA components. The rotation over the night totaled 74\textdegree, meaning each frame had several reference frames for PCA and subtraction. The ADI reduction recovers some of the thermal emission of the disk, albeit very faintly.

\subsection{Notable Features of AB Aur}

\begin{figure*}[]
    \centering
    \includegraphics[width=\linewidth]{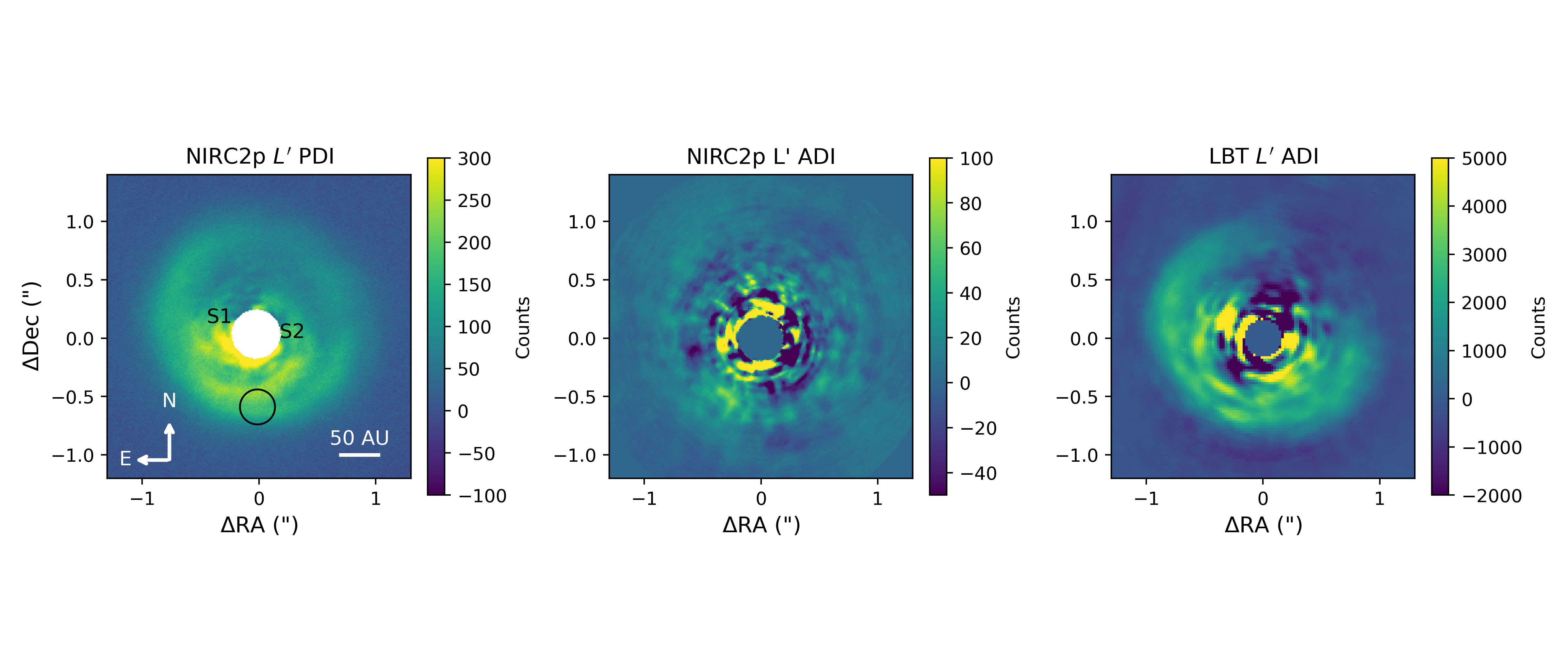}
    \caption{(Left) AB Aur in $L^\prime$ Stokes $Q_\phi$ with spiral arms S1 and S2 annotated, along with a circle centered on the coordinates of the candidate protoplanet from \citet{Currie2025ApJL}. (Middle) AB Aur in $L^\prime$ total intensity derived from the NIRC2 Polarimetry data with ADI. (Right) For comparison: AB Aur seen in total intensity $L^\prime$ with KLIP-RDI from \citet{betti2022detection}. It is clear that $L^\prime$ polarimetry is able to recover more detail of the disk's spiral structure and extended emission than was possible with traditional PSF subtraction methods on total intensity images.}
    \label{fig:spirals}
\end{figure*}

AB Aurigae is one of the nearest ($d$=163$\pm$2 pc, \citet{brown2018gaia}) and brightest Herbig Ae stars ($\sim$2-4 Myr old), and it hosts a well-studied transition disk with spiral arms and a potential embedded protoplanet \citep{boccaletti2020possible,Currie2022NatAs,Bowler2025AJ}. Spiral arms can be generated during the planet formation process as a protoplanet interacts with the disk \citep{bae2018planet}. The AB Aur disk has been imaged in scattered light at multiple optical and NIR wavelengths, including with polarimetry, but this work presents the first $L^\prime$ polarimetric imaging of the system, shown in Figure \ref{fig:abaur}. By comparing the magnitude of the polarized intensity signal compared to the magnitude of the total intensity speckles, we find a factor of at least $\sim$10$^2$ in speckle suppression for polarimetric differential imaging without the use of a coronagraph as measured in this dataset.

In these $L^\prime$ polarimetric images of the system, we clearly recover the S1 and S2 inner spirals as identified in \textit{H}-band scattered-light imaging from \citet{boccaletti2020possible} (Figure \ref{fig:spirals}), also previously detected in near-infrared imaging from HiCIAO \citep{hashimoto2011direct} and ALMA CO imaging \citep{tang2017planet}. The new NIRC2-Pol images, thanks to polarimetric differential imaging, reveal significantly more structure than previous $L^\prime$ imaging with the Large Binocular Telescope (LBT), which were poorly recovered due to limitations of the angular differential imaging (ADI) and reference differential imaging (RDI) techniques employed \citep{Jorquera2022ApJ,betti2022detection}.

We see no evidence for enhanced polarimetric emission at the location of the protoplanet candidate in the new $L^\prime$ polarized intensity images. First detected in 2022 via Subaru/SCExAO/CHARIS and \textit{Hubble} imaging \citep{Currie2022NatAs}, available data on AB Aur b have been complicated. AB Aur b was not seen in $L^\prime$ observations with LBT \citep{Jorquera2022ApJ}, nor was it detected in Paschen $\beta$ imaging of the system \citep{Biddle2024AJ}. \citet{Zhou2023AJ} found that the UV-optical emission of the system was inconsistent with that expected for an accreting protoplanet, while H$\alpha$ observations with \textit{Hubble}/WFC3 were inconclusive \citep{Zhou2022ApJL,Bowler2025AJ} and those from VLT/MUSE were interpreted as evidence of accretion onto AB Aur b \citep{Currie2025ApJL}. The lack of detection in the images presented here, however, does not indicate the absence of a protoplanet. 
A more thorough investigation of the contrast and planet detection limits in these images is beyond the scope of this brief letter, and is left for future work.

\vspace{2mm}
\section{Conclusions}

NIRC2 Polarimetry (NIRC2-Pol) enables infrared polarimetric imaging and low-resolution spectropolarimetry on Keck II across \textit{J} through $L^\prime$ bands. NIRC2-Pol is unique, as Keck II is the largest telescope (10 m) on which infrared polarimetry capabilities are available, and the only that can be combined with a vortex coronagraph. The new mode has been rigorously tested in both daytime and on-sky commissioning observations, and is now available to the community. Quantitative calibration of the system's instrumental polarization is underway \citep{zhang2026inprep} and a data processing pipeline is in development to aid observers \citep{lewis2026inprep}; estimates show the instrumental polarization is on the order of 1\% in $L^\prime$. The first on-sky results with NIRC2-Pol provide the first $L^\prime$ polarimetry of AB Aur's disk and the clearest $L^\prime$ imaging of its spiral features to date, illustrating the potential of this new mode; this data set should be further explored with future modeling work, including an investigation of the putative AB Aur protoplanet.

\section*{acknowledgments}
This material is based upon work supported by the National Science Foundation Astronomy \& Astrophysics Postdoctoral Fellowship Award No. 2401654 for author BLL. Any opinions, findings, and conclusions or recommendations expressed in this material are those of the authors(s) and do not necessarily reflect the views of the National Science Foundation. This work was also supported by the Mt. Cuba Astronomical Foundation and the University of California Observatories Mini-Grant Program. Thank you to Sarah Betti for providing the files for the LBT AB Aur imaging for comparison. J.N.A was supported by NASA through the NASA Hubble Fellowship grant \#HST-HF2-51547.001-A awarded by the Space Telescope Science Institute, which is operated by the Association of Universities for Research in Astronomy. JL, CAC, and MF acknowledge support from the Heising-Simons Foundation under grant No. 2022-354 and from the National Science Foundation under grants No. 1909641.

Some of the data presented herein were obtained at Keck Observatory, which is a private 501(c)3 non-profit organization operated as a scientific partnership among the California Institute of Technology, the University of California, and the National Aeronautics and Space Administration. The Observatory was made possible by the generous financial support of the W. M. Keck Foundation. The authors wish to recognize and acknowledge the very significant cultural role and reverence that the summit of Maunakea has always had within the Native Hawaiian community. We are most fortunate to have the opportunity to conduct observations from this mountain.
\section*{acknowledgments}

%

\vspace{5mm}
\facilities{Keck II (NIRC2)}


\software{NumPy \citep{numpy},
IPython \citep{ipython}, 
Jupyter Notebooks \citep{jupyter},
Matplotlib \citep{matplotlib}, 
Astropy \citep{astropy:2013, astropy:2018}, 
SciPy \citep{scipy}}
ADI.jl \citep{Lucas2020}




\bibliography{nirc2pol}
\bibliographystyle{aasjournal}



\end{document}